\documentstyle[sprocl]{article}

\input{psfig}
\def\NPB{{\em Nucl. Phys.} B}
\def\PLB{{\em Phys. Lett.}  B}

\def\PRD{{\em Phys. Rev.} D}
\def\ZPC{{\em Z. Phys.} C}
\def\be{\begin{equation}}
\def\ee{\end{equation}}
\def\bea{\begin{eqnarray}}
\def\eea{\end{eqnarray}}
\newcommand{\z}{&&\hspace*{-1cm}}

\def \ms {{\overline{\mbox{MS}}}}

\begin{document}

\title{QCD RELATIONS BETWEEN STRUCTURE FUNCTIONS AT SMALL X}

\author{ A.C. Kotikov}

\address{
Particle Physics Laboratory\\ Joint Institute for Nuclear Research\\
141980 Dubna (Moscow Region), Russia}

\author{ G. Parente }

\address{Departamento de F\'\i sica de Part\'\i culas \\
Universidade de Santiago de Compostela\\
15706 Santiago de Compostela, Spain}


\maketitle\abstracts{
We present very simple non-integral relations between deep inelastic
structure functions $F_L$, $F_2$ and the gluon distribution
at small $x$ based on perturbative QCD which are useful for
the phenomenological analysis of data at low $x$.
As an application we extract the deep inelastic scattering cross-sections
ratio $R= \sigma _L/\sigma _T$
in the range $10^{-4} \leq x \leq  10^{-2}$
from $F_2$ HERA data.
}

\section{Introduction}

For experimental studies of high energy hadron-hadron and lepton-hadron
processes it is necessary to know in detail the
values of the parton (quark and gluon) distributions
(PD) of nucleons, especially at small values of $x$. 
Of great relevance is the
determination of the gluon density at low x, where gluons
are expected to be dominant.

The basic
information on the gluon
structure of nucleons is extracted from the measurement of
the deep inelastic structure function $F_2$ in
lepton-hadron scattering (DIS). One of the usual procedures
compares experimental data with the theoretical
prediction for $F_2$ obtained from the solution of a system
of complicated
coupled integro-differential quark and gluon evolution equations.
It is also possible to extract the gluon distribution more directly
from $F_2$ scaling violations using
a very simple relation with the $Q^2$ derivative of $F_2$
\cite{PRYTZ}$^{\!-\,}$\cite{DUCATI}.

By other part, $F_L$ or the ratio $R=F_L/(F_2 - F_L)$, 
is also an interesting quantity, because it is a very sensitive QCD
characteristic.
For examplle, future $F_L$ measurements \footnote{At present there
are only preliminary measurements by H1 \cite{H1FL}},
will be used as a signal of the gluon structure at low 
$x$ \cite{COOPER}.

In perturbative QCD, there is the possibility to connect
$F_L$ with $F_2$ due to the fact that at small $x$
the DIS structure functions depend really on only
two {\it independent} functions,
the gluon and the singlet quark distribution (the nonsinglet quark
density is negligible at small $x$), which in turn
can be expressed in terms of $F_2$ and its derivative
$dF_2/dlnQ^2$.

In this article
we present very simple linear relations \footnote{Some of them 
have been already published \cite{KOPA,KOPAFL,KOPAR,KOTIJETP95}.} 
between the gluon density and $F_L(x,Q^2)$ with $F_2(x,Q^2)$
and $dF_2(x,Q^2)/dlnQ^2$ at small $x$. 
Using these formulas we exploit the possibility of extracting information
about the gluon distribution and $F_L$ at small $x$,
directly from the measurement of the $F_2$ scaling violations.
This method complement the standard analysis where quarks
and gluons, determined from complex fits to data,
are integrated for the calculation of $F_L$. With our formulas
it is possible to take into account the experimental uncertainty in
the theoretical calculation more directly.

The standard intial form for
the singlet quark
$s(x,Q^2_0)$ and gluon $g(x,Q^2_0)$ distributions  
\footnote{We use PD multiplied by $x$
 and neglect the nonsinglet quark distribution at small $x$.}
at some $Q^2_0$ are parameterized by \cite{4}:
\begin{eqnarray} 
p(x,Q^2_0) & = & A_p
x^{-\delta_p} (1-x)^{\nu_p} (1+\epsilon_p \sqrt{x} + \gamma_p x)
~~~~(p=S,g)
\label{1} 
\end{eqnarray}
Until the recent time the value of $\delta_p$
was a matter of discussion \cite{9}$^{\!-\,}$\cite{7},
but the new HERA data
\cite{F2H1,ZEUSGLU95} start to overcome this controversy.  
From the theoretical side, the type of evolution of the PD
in Eq. \ref{1} depends on the value and form of
$\delta_p$. For example \cite{Mar,KOTIYF93,EKL}, a $Q^2$-independent
$\delta_S = \delta_g$ obey the
DGLAP equation  when $x^{-\delta_p} \gg 1$. However, if $\delta_p (Q^2_0) =0$
in some point $Q^2_0 \geq 1 GeV^2$, the behaviour $p(x,Q^2) \sim Const$
is not compatible with DGLAP evolution and a more singular
behaviour is generated \cite{7,BF,KOTILOWX}.
These cases have been recently evolved to a common picture \cite{KOTILOWX}  
where partons are really a combination of
two solutions (at $x^{-\delta_p} \gg 1$ and at $\delta_p \sim 0$) linked
at some $Q^2$ point. For a Regge-like
form of structure functions, one obtains \cite{KOTILOWX}
$ p(x,Q^2) \sim x^{-\overline \delta_p(Q^2)} $
with next-to-leading order (NLO) $\overline \delta_S(Q^2) \neq
\overline \delta_g(Q^2)$ intercept trajectories.
 Without any restrictions, it generates the double-logarithmical behaviour,
\begin{eqnarray}
 p(x,Q^2) \sim \exp{\biggl(2
   \sqrt{\overline \delta_p(Q^2)ln\frac{1}{x}}\biggr)}  
\label{2} \end{eqnarray}
where at NLO and for $f=4$ active quarks one has \cite{KOTILOWX}:
$$\overline \delta_g(Q^2)~=~ \frac{36}{25} t -\frac{91096}{5625} r,~~
\overline \delta_S(Q^2)~=~\overline \delta_g(Q^2)  -20 r$$
being $t~=~ln(\alpha(Q^2_0)/\alpha(Q^2))$ and $r~=~
\alpha(Q^2_0) - \alpha(Q^2)$ \footnote{Hereafter
we use $ \alpha(Q^2)= \alpha_s(Q^2)/{4 \pi}$.
Sometimes it is denoted as $ \alpha $ to shorten long formulae.}.

\section{The gluon and $F_L$ as a function of $F_2$ and
$dF_2/dlnQ^2$}

By lack of space we will only present the final
formulas of the calculation while the details will be given
elsewhere \cite{INPREP}. 
Assuming the {\it Regge-like behaviour} of Eq. \ref{1} for
$x^{-\delta_p} \gg 1$, one can replace at small $x$
the convolution integrals
by ordinary products in 
the $Q^2$ evolution equations and in
the formulas relating $F_2$ and $F_L$ to PD. 
From the $F_2$ equation one can extract the
singlet quark combination $S(x,Q^2)$
as a function of $F_2(x,Q^2)$ and $g(x,Q^2)$ and substitute
it into the equations for $F_L$ and for $dF_2/dlnQ^2$. 

The case of the non-Regge type behaviour has to be treated independently,
but it is possible to combine both cases (Regge and non-Regge types)
in a single formula valid for any value of $\delta_p$:
\begin{eqnarray} \z
g(x, Q^2)  = - \frac{2f}{\alpha e}
\frac{ 1}{ \gamma^{(0),1+\delta_g}_{sg} + 
\tilde \gamma^{(1),1+\delta_g}_{sg} \alpha }  \biggl[ 
\frac{d F_2(x, Q^2)}{dlnQ^2} 
+ \frac{\alpha }{2} 
 \gamma^{(0),1+\delta_S}_{ss} 
F_2(x,Q^2) \nonumber \\ \z
~~~~~~~~~~~~ ~+~ O(\alpha^2, x^{1-\delta})  \biggr],  
\label{14.9}  \\
\z  F_L(x, Q^2)  = 
\alpha 
  B^{S,\delta_S}_{L} 
F_2(x,Q^2)- 2 
\frac{B^{g,1+\delta_g}_L \Bigl(1+ \alpha \tilde R^{g,1+\delta_g}_L \Bigr)}{ 
\gamma^{(0),1+\delta_g}_{Sg} + 
\tilde \gamma^{(1),1+\delta_g}_{Sg} \alpha } 
\biggl[
\frac{d F_2(x, Q^2)}{dlnQ^2} \nonumber \\ \z
~~~~~~~~~~~~ ~+~ \frac{\alpha }{2}  
 F_2(x,Q^2) \biggr]
+
O(\alpha^2, x^{1-\delta_p}), 
\label{15.1} 
\end{eqnarray}
where $e = \sum_i^f e^2_i$ is the sum of squares of $f$ quark charges.
The variables $B^{l,\eta}_k$ ($k=2,L$) and $\gamma^{(0),\eta}_{pl}$ 
$(p,l=g,S)$
are respectively the one loop parts of the Wilson coefficients and anomalous
dimensions of the operators. The second order quantities
are related to the two loop
Wilson coefficients and anomalous dimensions by:
\begin{eqnarray} \z
\overline R^{g}_{L} ~=~  R^{g,1+\delta_g}_{L} - B_2^{g,1+\delta_g}
 B^{S,1+\delta_S}_{L} / B^{g,1+\delta_g}_{L},  \nonumber \\ \z
\overline    \gamma^{(1)}_{Sg} ~=~
   \gamma^{(1),1+\delta_g}_{Sg} + B_2^{S,\delta_S} 
\gamma^{(0),1+\delta_g}_{Sg} + B_2^{g,1+\delta_g} \bigl(2\beta_0 + 
  \gamma^{(0),1+\delta_g}_{gg} - \gamma^{(0),1+\delta_S}_{SS} \bigr)
 \nonumber  
 \end{eqnarray} 

All these coefficients have to be continued analytically from the
well known integer values $n$ to the non-integer ones $\eta$.

The values of
$\tilde \gamma^{(1)}_{Sg}$ and $\tilde R^{g}_{L}$
 coincide with $\overline \gamma^{(1)}_{Sg}$ and 
$\overline R^{g}_{L}$, respectively, with the replacement,
\begin{eqnarray}
\frac{1}{\delta_g} \to \int^1_x \frac{dy}{y}~ \frac{g(y,Q^2)}{g(x,Q^2)} 
\label{14.1} 
\end{eqnarray}
In the cases $x^{-\delta_g} \gg Const$ and $\delta_g \to 0$, the r.h.s. of 
Eq. \ref{14.1} leads to $1/\delta_g $ and $1/\tilde \delta_g $,
respectively, where
\begin{eqnarray}
\frac{1}{\tilde \delta_p} = 
\sqrt{\frac{ln(1/x)}{\overline \delta_p(Q^2)}} 
- \frac{1}{4\overline \delta_p(Q^2)} 
\biggl[1+
\frac{1}{8\sqrt{4\overline \delta_p(Q^2)\ln(1/x)}} +
 O\Bigl(\frac{1}{4\overline \delta_p\ln(1/x)} \Bigr) \biggr]
\! \label{14} 
\end{eqnarray}
The comparison of our results with others is performed elsewhere
\cite{INPREP}.

\section{The extraction of the ratio $R$ at low $x$} \indent

As an application we have extracted
the ratio $R(x,Q^2)$ from H1 1994
data \cite{F2H1}, determining the slopes dF$_2$/dlnQ$^2$
from straight line fits \cite{H1PREP95,ZEUSGLU95}.
The systematic errors have been taken from an early
analysis of H1 \cite{H1PREP95}.
The running coupling constant
$\alpha_s(Q^2)$ has been calculated at two loops using
$\Lambda^{(4)}(\ms)=225 MeV$.
 
Figure 1 shows the extracted ratio $R$ at
$Q^2 = 20$ GeV$^2$ for $\delta_S=0.3$ and  
two different exponent for the gluon,
$\delta_g = \delta_S$ and 
 $\delta_g = \delta_S +0.05$.
These values are very close to those obtained
by various groups from QCD fits to H1 data \cite{F2H1,MRS96}.
Fig. 1 also shows some experimental data points \cite{FLDATA}
at high $x$.
\begin{figure}
%\rule{5cm}{0.2mm}\hfill\rule{5cm}{0.2mm}
%\vskip 6.cm
%\rule{5cm}{0.2mm}\hfill\rule{5cm}{0.2mm}
\psfig{figure=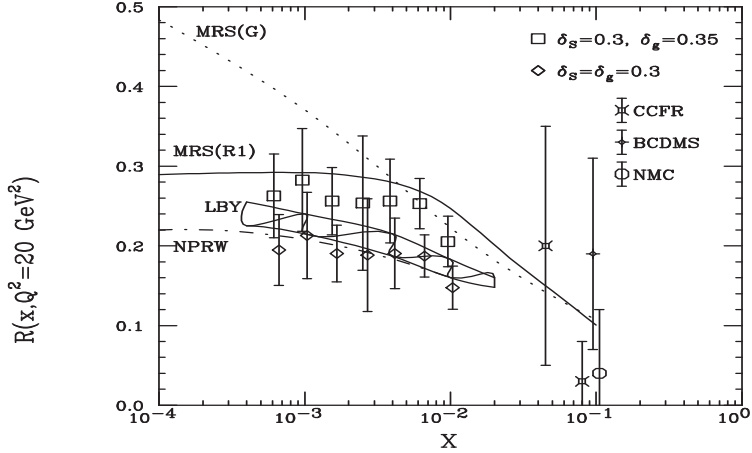,height=6cm,width=10cm}
\caption{The ratio $R$ extracted from H1 data. The lines are the
predictions from different models (see text). The band represent
the uncertainty from a DGLAP analysis of HERA data}
\end{figure}
For comparison we have also plotted various predictions for $R$ using
QCD formulae at O($\alpha_s^2$)
\footnote{The quark singlet and non-singlet kernels in $\ms$ scheme
are taken from
ref. \cite{FLNLO}. It was noted in \cite{LARINVERMASEREN}
that the gluon kernel given in \cite{FLNLO} is erroneous.
We use the correct result given in \cite{VANNEERVEN} and \cite{KAKO}.}
and parton densities extracted from fits to HERA data.
The large difference between the result from MRS(G) and the latest
set MRS(R1) \cite{MRS96} shows the sensitivity of R
to the update of these parton densities to new HERA data.
One can notice that 
there is good agreement between MRS(R1) prediction and our points when
$\delta_g = \delta_S +0.05 =0.35$, even though it is not surprising
because these values were taken from the MRS analysis 
\cite{MRS96}. However our result gives extra information about
the uncertainty.
A more precise future measurements at low $x$
should lie within the error bars of the results presented in Fig. 1. 

By other part recent theoretical predictions
on $R$ based on conventional NLO DGLAP evolution analysis of HERA data
(LBY) \cite{LBY} and on the dipole picture of BFKL dynamics
(NPRW) \cite{NPRW}, both finding values $\delta_S \equiv \delta_g
\approx 0.3$, lie between the two of our above cases.
%
%------------   FIN -------------
%
\section*{Acknowledgments}

This work was partially supported by CICYT (AEN96-1773), by
Xunta de Galicia (XUGA-20604A96), and by RFFR (95-02-04314-a).

\end{document}